# ENERGY AWARE ROUTING SCHEME FOR MOBILE AD HOC NETWORK USING VARIABLE RANGE TRANSMISSION

Pinki Nayak[#1], Rekha Agrawal [$1], and Seema Verma[*3]

[1]Department of Electronics & Communication, Amity School of Engg. & Technology,
GGSIP University, New Delhi-110061, INDIA
[#]`pinki_dua@yahoo.com`
[$]`rarun96@yahoo.com`
[2]AIM & ACT, Banasthali University
Rajasthan-304022, INDIA
`seemaverma3@yahoo.com`

## ABSTRACT

*A Mobile Ad hoc Network is a collection of mobile nodes that dynamically forms networks temporarily without the need for any pre-existing infrastructure. Today, one of the main issues in MANETs is the development of energy efficient protocols due to limited bandwidth and battery life. The nodes in MANETs operate by a battery source which has limited energy reservoir. Power failure of a node affects the node's ability to forward packets on behalf of others, thus reducing the network lifetime. The conventional MANET routing protocols s. a. DSR and AODV use common transmission range for transfer of data and does not consider energy status of nodes. This paper discusses a new energy aware routing (EAR) scheme which uses variable transmission range. The protocol has been incorporated along with the route discovery procedure of AODV as a case study. Both the protocols are simulated using Network Simulator-2 and comparisons are made to analyze their performance based on energy consumption, network lifetime and number of alive nodes metrics for different network scenarios. The results show that EAR makes effective node energy utilization.*

## 1. INTRODUCTION

A Mobile Ad hoc *N*etwork (MANET) is a network formed without any central administration. It consists of nodes that use a wireless interface to send & receive packet data. These nodes in the network are mobile and can serve as routers and hosts, thus can forward packets on behalf of other nodes and run user applications. This allows people and devices to seamlessly inter-network in areas with no pre-existing communication infrastructure. Significant examples of MANET include establishing survivable dynamic communications for emergency/rescue operations, disaster relief efforts, military networks, business indoor application, home intelligence devices [1].

Developing routing protocols for MANETs has been a challenging task because of its dynamic topology, bandwidth constrained wireless links and resource (energy) constrained nodes. Many proactive and reactive protocols have been proposed which try to satisfy various properties, like:

---

[1] Pinki Nayak is also a research scholar at Banasthali University, Banasthali, Rajasthan, India

efficient utilization of bandwidth, battery capacity, fast route convergence, optimization of metrics (like throughput and end-to-end delay), and freedom from loops.

This paper tries to address the problem of energy efficient routing to increase the lifetime of the network. Mobile hosts, today are powered by battery, therefore efficient utilization of battery energy becomes very important. The energy resources of actively participating nodes get depleted faster than other nodes, which in some cases, may lead to partitioning of the network, thus decreasing the lifetime of the network. For this reason, reducing the energy consumption is an important issue in ad hoc wireless networks [2]. The three major ways of increasing the life of a node are efficient battery management, transmission power management and system power management. In this paper, a scheme has been proposed to minimize the energy consumption at the nodes, thus maximizing the network lifetime. Transmission power control approach is used to adjust the power levels at node. Common power levels are used during Route Discovery. New power levels are calculated between every pair of nodes based on distance.

The rest of the paper is organized as follows. Section 2 presents a discussion on the related work in energy aware routing protocols. Section 3 gives a brief description of the existing routing protocols. In section 4, the detailed working of the proposed EAR is discussed. Section 5 includes the simulation environment setup used in NS-2 simulator. The simulation results are explained in section 6. Finally, section 7 gives conclusions.

## 2. RELATED WORK

Many research efforts have been devoted for developing energy efficient routing algorithms. Node's energy is minimized not only during active communication but also when they are in inactive state. Transmission power control and load distribution are two approaches used to minimize the active communication energy of individual nodes and sleep/power-down mode to minimize energy of nodes during inactivity. In transmission power control approach choosing a high transmission power reduces the number of forwarding nodes needed to reach the required destination, but creates excessive interference in a medium that is commonly shared whereas, choosing a lower transmission power reduces the interference seen by potential transmitters but packets require more forwarding nodes to reach their required destination. The specific goal of the load distribution approach [13-15] is to balance the energy usage of all mobile nodes by selecting a route with underutilized nodes rather than selecting the shortest route [5, 6, 8].

Some research proposals based on transmission power control are discussed in [3, 4]. Flow Augmentation Routing (FAR) [9] finds the optimal routing path in a static network, for a given source–destination pair that minimizes the sum of link costs along the path. Online Max-Min Routing (OMM) [10] for wireless ad-hoc networks optimizes the lifetime of the network as well as the lifetime of individual nodes by maximizing the minimal residual power, which helps to prevent the occurrence of overloaded nodes. Power-aware Localized Routing (PLR) [11] is a localized, fully distributed energy-aware routing algorithm, with the assumption that a source node has the location information of its neighbors and the destination. The main goal of Minimum Energy Routing (MER) [12] is not to provide energy efficient paths but to make the given path energy efficient by adjusting the transmission power just enough to reach to the next hop node. The authors in [3] investigates the impact of variable range power control on physical layer and network layer connectivity and shows that variable range increases network lifetime over common range transmission. An optimal transmission range for nodes during flooding the route request messages is given in [4].

## 3. CONVENTIONAL ROUTING PROTOCOLS

Routing in MANET depends on factors like topology, selection of routers, location of request initiator etc. in finding the path quickly and efficiently. The traditional routing algorithms are classified as Proactive and Reactive protocols. These algorithms lack energy awareness of the nodes in the network Reactive routing protocols discover or maintain a route as needed. This reduces overhead that is created by proactive protocols. Reactive routing protocols can be classified as source routing and hop by hop routing. Dynamic Source Routing (DSR) is an example of source routing, whereas Ad-hoc On Demand Distance Vector Routing (AODV) is a hop by hop routing protocol.

DSR (Dynamic Source Routing) [17] is an on-demand, simple and efficient routing protocol for multi-hop wireless ad hoc networks of mobile nodes. DSR uses source routing whereby all the routing information is maintained (continuously updated) at mobile nodes instead of relying on the routing table at each intermediate device. The protocol is composed of two main mechanisms-'Route Discovery' and 'Route Maintenance', which works together entirely, on-demand. The protocol allows multiple routes to destination, loop-free routing, support for unidirectional links, use of only 'soft state' in routing and rapid discovery when routes in the network change.

AODV (Ad hoc on-demand distance vector) [18] is a dynamic, self-starting, loop free, multi-hop on-demand routing for mobile wireless ad hoc networks. AODV discovers paths without source routing and maintains route table instead of route cache. It allows mobile nodes to respond to link breakages, changes in network topology in a timely manner through Route Error (RERR), Route Request (RREQ) and Route Reply (RREP) messages. It maintains active routes only while they are in use using sequence numbers and delete unused routes (stale).

## 4. PROPOSED ALGORITHM

The main aim of the proposed algorithm is to make the network energy aware. Energy efficient design of the protocol is generated by varying the transmission range of the nodes. Variable transmission range means controlling the power level for each packet in a distributed manner at each node, thus affecting energy consumption of the network. Choosing a higher transmission range reduces the number of nodes needed to reach the destination but creates large interference, whereas reducing the transmission range demands more number of forwarding nodes leading to less energy utilization. Each node communicates with the neighboring nodes during Route Discovery phase. Once the route is known, each individual node then controls the transmission range as per the distance between source and destination node, so that optimum energy is utilized for packet transmission. The proposed algorithm is explained below:

The steps involved in EAR are:

**Step 1**: When any node needs to send data, it generates the Route Request (RREQ) packet and broadcast it to its neighbor with initial common transmission range of 250m.

**Step 2**: The route reply messages from the intermediate nodes contain two fields' *locX* and *locY* that stores the location of the node sending the route reply.

**Step 3**: In AODV, the path is established for the first RREP received. But in EA-AODV, each node waits for a time ($T\_wait$) till it receives all the RREP messages destined for the node.

**Step 4**: The node then calculates the distances between the nodes from where the RREP message is received and itself. This is done using own location and the locations of the intermediates nodes.

**Step 5**:   Now, the node with minimum distance is selected calculated in step 4 and its location is also updated in the routing table as two entries *n_hopX* and *n_hopY*.

**Step 6**: The transmission energy is then calculated using the Friis transmission equation in free space (Appendix) based on the distance, which is the distance between the current node and the next hop node in the said algorithm. The received power threshold of the node is kept constant throughout.

**Step 7**:   The route between source and destination is maintained for data transfer.

**Step 8**:    If the route is broken, repeat from step 1.

## 5. SIMULATION SETUP

The simulations are carried out using the event driven simulation tool Network Simulators-2 [2] (ver. 2.34) and the wireless extensions provided by CMU [3]. The used channel is Wireless channel/Wireless Physical, Propagation model is Free Space Propagation Model, Queuing model is Drop Tail/Priority Queue, Mobility model is Random Waypoint model and MAC protocol is 802.11.

The simulation setup consists of an area of 1000 X 1000 $m^2$ with different number of nodes for each simulation. To emulate the dynamic environment, all the nodes move around in the entire region. Varying speeds with minimum 2m/sec and maximum of 40 m/sec have been considered. Constant Bit Rate (CBR) traffic source with packet size of 512 bytes is taken. Different source-destination pair (5 – 25 connections) was used to establish the routes. All the simulations were run for a period of 200 sec. The initial energy of each node was set as 100 Joules with transmission and reception power of 5 W and 1W respectively. Table 2 gives the simulation parameters used.

Table 2. Simulation Parameters

| Parameter | Value |
| --- | --- |
| Number of Nodes | 50 |
| Grid Area | 1000m x 1000m |
| No. of Connections | 50% of number of nodes |
| Pause Time | 0 sec |
| Speed | 5 m/s, 10m/s, 40m/s |
| Traffic Model | CBR |
| Data Packet Size | 512 bytes |
| Data Packet Interval | 4 packets/sec |
| Simulation Time | 200 sec |
| Initial Energy of Node | 100 J |
| Transmitted Power | 5 W |
| Received Power | 1W |
| Idle State Power | 0.0005 W |
| Sleep State Power | 0.0002 W |
| Transition from sleep to active state | .03W |

## 6. RESULTS AND DISCUSSION

As a case study, EAR has been implemented on AODV. Simulations have been conducted to compare the performances of AODV and EAR protocols. Performance comparisons have been made on the following parameters

(i) **Network Lifetime**: The network lifetime can be defined as [7]

- the time taken for k nodes to die
- the time taken for the first node to die
- the time taken for all the nodes in the network to die

In this paper, the first definition is adopted.

(ii) **Total Energy Consumption**: This metric gives the energy consumption in the network due to routing packets. It is given as

$E_{Tx} = (1.65 * \text{Packet Size})/2 \times 10^6$ \hfill (1)

$E_{Rx} = (1.1 * \text{Packet Size})/ 2 \times 10^6$ \hfill (2)

where, $E_{Tx}$, $E_{Rx}$, are the transmitted and received energy of a node utilized in routing a packet.

(iii) **Number of Alive Nodes (AN)**: This efficiency metric gives the status of the number of nodes left with more than 50% of their energy at the end of simulation.

### 6.1. Network Lifetime

Network Lifetime is a vital performance metric while comparing routing protocols. First, the network lifetime is observed through number of exhausted nodes. Thus, the lesser the number of energy exhausted nodes; the better is the network lifetime. Later network lifetime is computed as the time when 50% of the total nodes die out.

The comparison of network lifetime between AODV and EAR with nodes moving at speed of 10 m/sec and 40 m/sec respectively is shown in Figure 1. Number of connections is taken as 10. It is observed that energy consumption of participating nodes in the network starts increasing from 91 sec of the simulation time for AODV and from 96 sec for EAR when the speed of nodes is 10m/sec. As the speed is increased to 40m/sec, EAR still shows a better distributed network lifetime as compared to AODV. This is because, as the transmitted power is adjusted according to the shortest distance in EAR, the available node energy is effectively used decreasing the number of energy exhausted nodes.

Figure 2 gives the network lifetime with varying pause time. EAR shows a better performance as compared to AODV at lower values of pause time. With increase in pause time the network becomes less mobile and EAR behaves like a normal AODV scheme

## 6.2. Total Energy Consumption

Figure 3 shows the total energy consumed in the network with varying number of nodes. The number of connections is taken as 50% of the number of nodes in the network and the nodes are moving at speed of 10m/sec with pause time of 0 seconds. Initially, with lesser number of nodes energy consumption in EAR and AODV is comparable. As the network becomes denser, energy consumption for EAR is less than AODV. This is because the variable transmit power keeps the total energy consumption lower in EAR.

Figure 4 gives the variation of total energy consumed with number of connections in a network with 50 nodes. It is clearly seen that EAR outperforms AODV as the load in the network increases. The energy consumption in the network is reduced by 10%-20% in EAR in comparison with AODV protocol

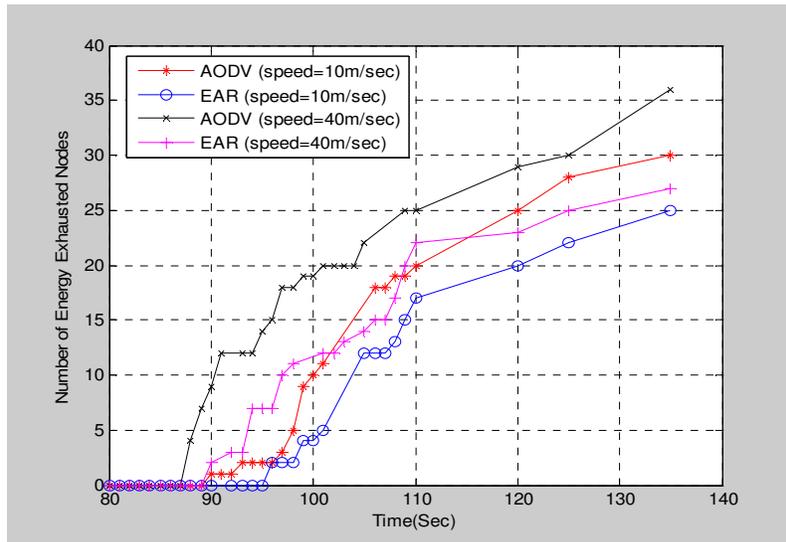

Figure 1. Network Lifetime for Speed 10m/sec and 40m/sec

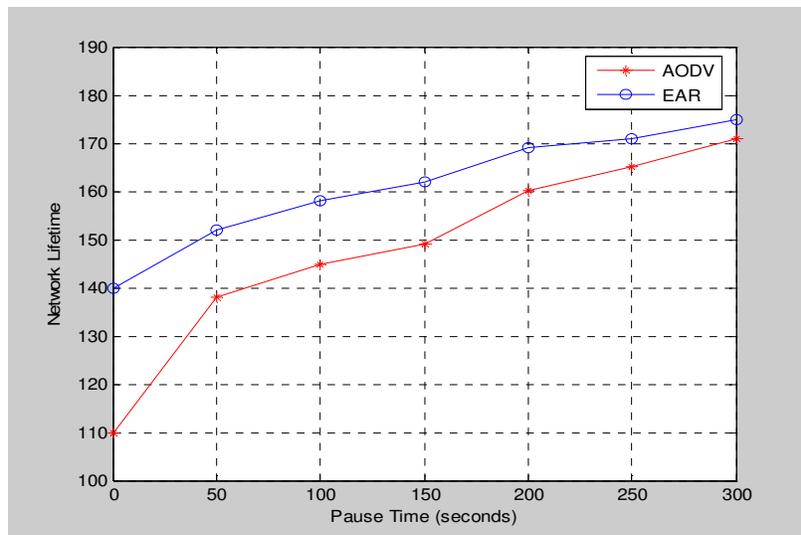

Figure 2. Network Lifetime with Pause Time

International Journal of Ad hoc, Sensor & Ubiquitous Computing (IJASUC) Vol.3, No.4, August 2012

In Figure 5 energy consumption is varied with speed of the nodes. At higher speed, the nodes are more mobile causing frequent path breaks, which requires new route discovery, increasing routing overhead. These causes increase in energy consumption as compared to low speeds. Again EAR performance is 10%-20% better than AODV protocol.

### 6.3. Number of Alive Nodes (AN)

For Figure 6, simulations are carried for 50 nodes at speed of 10m/sec. Variation of **AN** with varying number of connections is plotted. Pause time of 0 seconds is considered. The greater the value of this metric the better is the protocol. As seen, number of alive nodes decreases with number of connections as number of nodes transmitting data i.e. the network load is increased. EAR shows better variation over AODV. The alive nodes left in EAR are more as compared to AODV protocol.

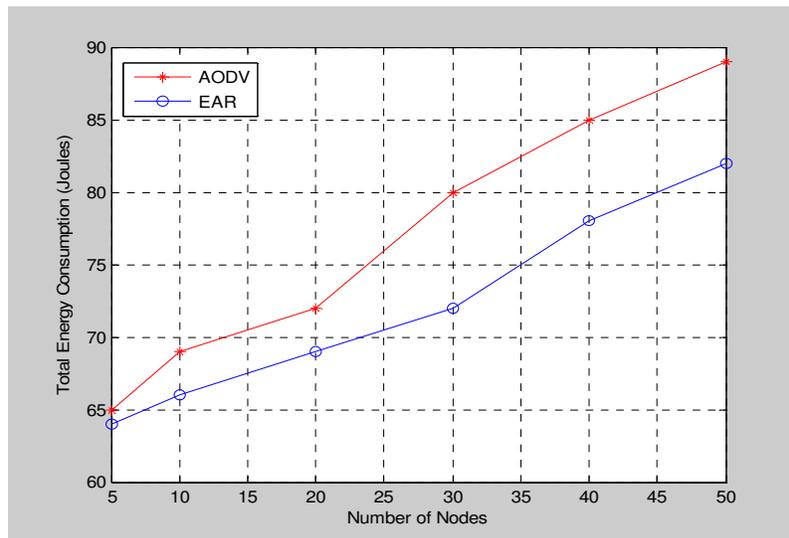

Figure 3. Total Energy Consumption with Number of Nodes

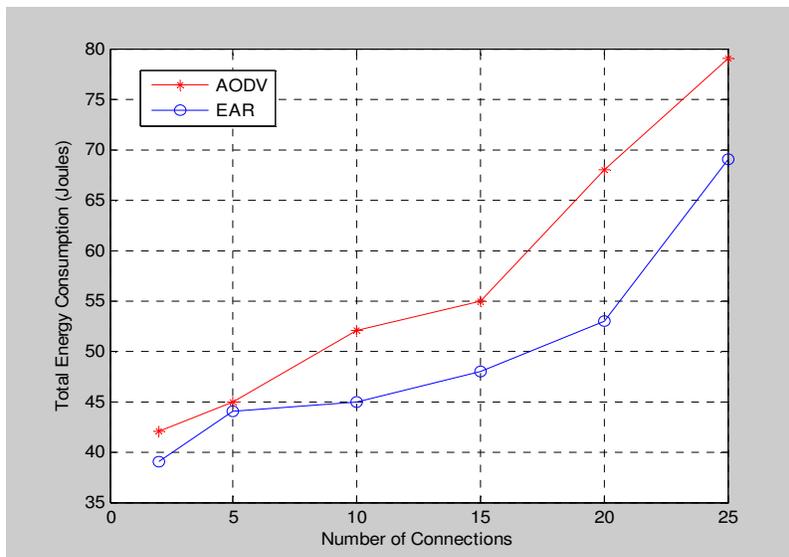

Figure 4. Total Energy Consumption with Number of Connections

International Journal of Ad hoc, Sensor & Ubiquitous Computing (IJASUC) Vol.3, No.4, August 2012

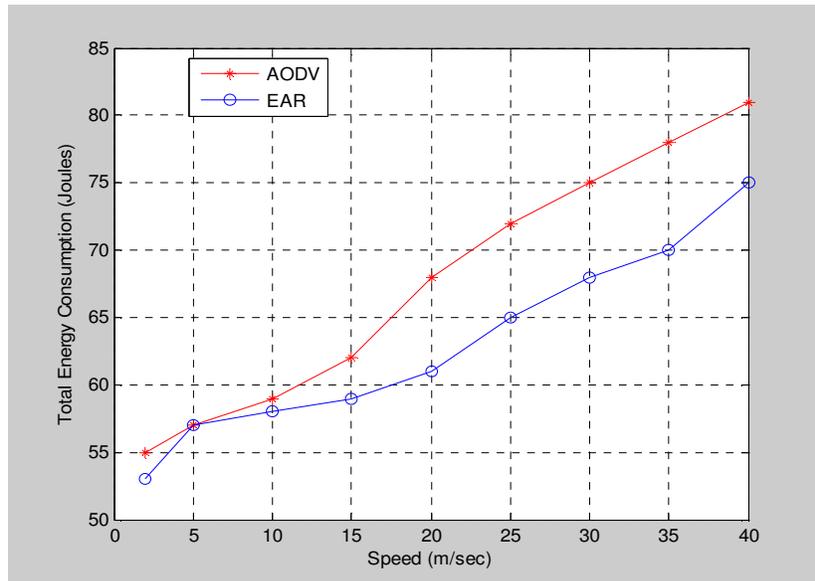

Figure 5. Total Energy Consumption with Speed

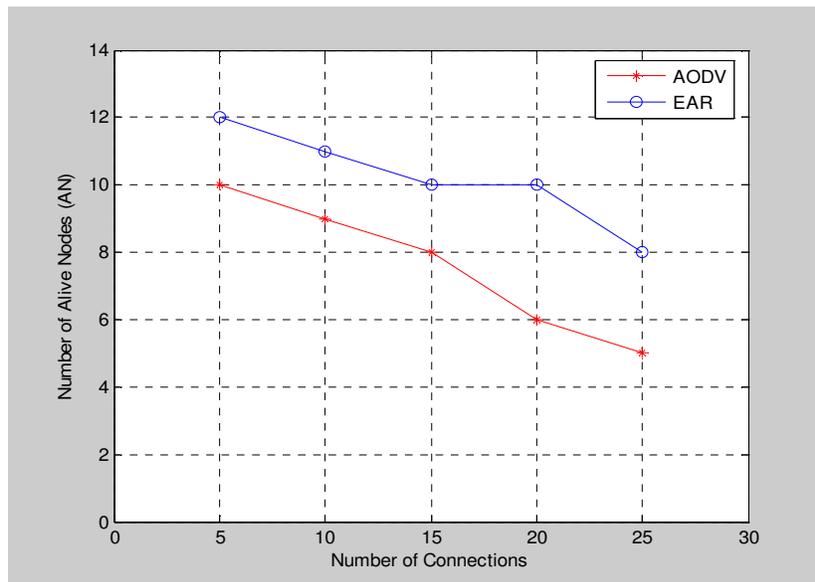

Figure 6. Number of Alive Nodes with Number of Connections

## 7. CONCLUSION

To design an energy efficient protocol has been one of the main challenges in MANETs. The proposed work aims at designing an efficient energy aware routing scheme for MANETs taking variable range transmission into consideration. Simulation results show that EAR shows superior performance as compared to common range AODV in terms of energy consumption and improves network lifetime. An average of 10%-20% increase in network lifetime and a 10% reduction in total energy consumption is observed. Number of alive nodes left at the end of the simulation is also increased by 10% in EAR as compared to AODV.

Each node's energy state has a big influence on the entire network lifetime. At the time of route selection, EA-AODV takes care of the distance between the nodes choosing nodes with least distance thereby decreasing the energy required to transmit data. Energy consumption at each node is thus improved using variable range transmission in proposed scheme. This is due to the transmitter power adjustments done at each node before transmitting the data, which makes utilization of different nodes in the network effective. Future work includes simulation of the proposed scheme for sparse mediums and real life scenarios and also for other metrics like link layer overhead, pause time, path optimality etc.

## **APPENDIX: TRANSMITTER POWER DETAILS**

The transmit power of the node needs to be modified, to vary the transmission range. In the proposed algorithm, free space propagation has been chosen for simulation. Friis transmission equation is used to calculate the transmit power of nodes, given as:

$$P_r = P_t G_r G_t \left(\frac{\lambda}{4\pi R}\right)^2 \qquad (3)$$

where, $P_r$ & $P_t$ are received & transmit powers respectively, $G_t$ & $G_r$ are the transmitter & receiver antenna gain, R is the distance between the nodes & $\lambda$ is the wavelength.

For the simulations, $G_t$ & $G_r$ are taken as unity, $\lambda$ is taken as 0.125 m (at 2.4 GHz operating frequency). The constant received power $P_r$ has been chosen as -84dBm as IEEE 802.11 [16] MAC protocol specifies a received power range from -81.0 dBm to -110 dBm at transmission bandwidth of 2 Mbps [13]. Thus, based on the measured distance between nodes, $P_t$ can be calculated using equation 3. Table 2 gives some values of $P_t$ with changing R. As seen the transmitter power requirement increases with increase in distance between nodes.

Table 2. Transmit Power for different distances

| Distance (m) | $P_t$ (dBm) |
|---|---|
| 50 | -9.2 |
| 100 | -3.23 |
| 150 | 0.3 |
| 200 | 2.8 |
| 250 | 4.73 |
| 400 | 8.9 |

**Authors**

**Pinki Nayak** received her M.Tech (Information Technology) from Guru Gobind Singh Indraprastha University, New Delhi, India. At present, she is working as Assistant Professor in Department of ECE at Amity School of Engg. & Tech., New Delhi and is also working towards the Ph.D. degree at Banasthali University, Banasthali, India. Her research interest lies in the area of Mobile Ad-hoc networks, Wireless Communucation.

**Dr. Rekha Agarwal** is working as Professor at, Amity School of Engg & Tech., New Delhi. She graduated in Electronics and Communication Engg. from Jiwaji University, Gwalior. Later, she obtained her Master's degree from MREC, Jaipur and then Ph.D. degree from Guru Gobind Singh Indraprastha University, Delhi. She has been actively involved in teaching and research and has published many research papers. Her teaching and research interest include multiuser detection, coding techniques, Ad-hoc Networks.

**Dr. Seema Verma** obtained her Master and Ph.D degree in Electronics from Banasthali University in 1999 & 2003. She is working as Associate Professor [Electronics] at Banasthali University. She is an active research supervisor and has presented many papers in various international conferences. She has published many research papers in various journals of repute. Her research areas are Coding theory, TURBO Codes, Wireless sensor networks, Aircraft Ad-hoc networks, Network Security & VLSI Design.